\documentclass[conference]{IEEEtran}
\IEEEoverridecommandlockouts
\usepackage{cite}
\usepackage{amsmath,amssymb,amsfonts}
\usepackage{algorithmic}
\usepackage{graphicx}
\usepackage{textcomp}
\usepackage{xcolor}
\usepackage{amssymb}
\usepackage[colorlinks=true, citecolor=black, urlcolor=black, linkcolor=black]{hyperref}

\usepackage{mwe}                       

\def\BibTeX{{\rm B\kern-.05em{\sc i\kern-.025em b}\kern-.08em
    T\kern-.1667em\lower.7ex\hbox{E}\kern-.125emX}}

\usepackage{mathtools}
\newcommand{\systemname}{our application\xspace}

\begin{document}

\title{\textit{Who should I trust?}A Visual Analytics Approach for Comparing Net Load Forecasting Models\\
\thanks{Parts of this work were supported by the U.S. Department of Energy Solar Energy Technologies Office and the Office of Electricity Sensors Program, and performed jointly at the Pacific Northwest National Laboratory under Contract DE-AC05-76RL01830 and at the Lawrence Livermore National Laboratory under Contract DE-AC52-07NA27344).}
}


\author{\IEEEauthorblockN{Kaustav Bhattacharjee\IEEEauthorrefmark{1}, Soumya Kundu\IEEEauthorrefmark{2}, Indrasis Chakraborty\IEEEauthorrefmark{3}, and Aritra Dasgupta\IEEEauthorrefmark{1}}

\IEEEauthorblockA{\IEEEauthorrefmark{1}\textit{Department of Data Science}, New Jersey Institute of Technology, USA. Email: \{kb526,\,aritra.dasgupta\}@njit.edu} 
\IEEEauthorblockA{\IEEEauthorrefmark{2}\textit{Optimization and Control Group}, Pacific Northwest National Laboratory, USA. Email: soumya.kundu@pnnl.gov}
\IEEEauthorblockA{\IEEEauthorrefmark{3}\textit{Center for Applied Scientific Computing}, Lawrence Livermore National Laboratory, USA. Email: chakraborty3@llnl.gov} 
}

\maketitle

\begin{abstract}
Net load forecasting is crucial for energy planning and facilitating informed decision-making regarding trade and load distributions. However, evaluating forecasting models' performance against benchmark models remains challenging, thereby impeding experts' trust in the model's performance. 
In this context, there is a demand for technological interventions that allow scientists to compare models across various timeframes and solar penetration levels. 
This paper introduces a visual analytics-based application designed to compare the performance of deep-learning-based net load forecasting models with other models for probabilistic net load forecasting. 
This application employs carefully selected visual analytic interventions, enabling users to discern differences in model performance across different solar penetration levels, dataset resolutions, and hours of the day over multiple months. We also present observations made using our application through a case study, demonstrating the effectiveness of visualizations in aiding scientists in making informed decisions and enhancing trust in net load forecasting models.
\end{abstract}

\begin{IEEEkeywords}
visual analytics, AI/ML, net load forecasting
\end{IEEEkeywords}

\section{Introduction}
Comparing multiple computational models for their performance is crucial for enhancing trust in model outcomes because it provides a basis for evaluating the reliability and consistency of each model~\cite{bui2020comparing, naghibi2015comparative,lu2021performance}. By assessing how different models perform under various conditions and scenarios, stakeholders can better understand their strengths, weaknesses, and overall effectiveness. Such comparative analysis helps to identify the most suitable model for specific tasks or applications, thereby instilling confidence in the reliability of the chosen model. This becomes more important while predicting the net load of an electric grid, which is defined as the difference between total electricity demand and generation from behind-the-meter resources like solar and distributed generators, and is influenced by various factors such as weather conditions and time of day~\cite{campbell2024clustering,NetLoadDefinition}. 

Accurate net load forecasting enables grid operators, policymakers, and energy providers to make informed decisions regarding energy trade, load distribution, and resource allocation. However, the rise of solar energy generation sources in residential settings has significantly impacted the performance of traditional net load forecasting models, highlighting the need for robust time-series forecasting techniques~\cite{SolarEnergyPenetration}. Collaborating with scientists, we developed a deep-learning model that integrates variables such as temperature, humidity, apparent power, and solar irradiance to achieve strong predictive performance and resilience in the face of missing data~\cite{sen2022kpf}. However, the model's sensitivity to seasonal variations and noisy inputs underscores the need for further exploration with domain experts. Hence, we developed an interactive tool that empowers users to investigate the model's performance across diverse time periods and input scenarios~\cite{bhattacharjee2024forte}.

However, while we initially recognized the importance of comparing the model's performance with traditional models, feedback from domain experts further emphasized its significance in building trust in model outcomes. Visual analytics can play a pivotal role here, as evidenced by prior research demonstrating its importance in enhancing trust in machine learning models~\cite{chatzimparmpas2020state}.
This is exemplified by the interactive tool we developed, which enabled domain experts to extract valuable insights concerning the model's sensitivity toward temperature and humidity. Moreover, recent discourse, as highlighted in \cite{kandakatla2020towards}, emphasizes the critical role of visual analytics in fostering trust-augmented applications of artificial intelligence and machine learning (AI/ML) within the energy sector. Building upon this, we designed  \systemname, incorporating carefully selected visual analytic interventions. These interventions facilitate the comparison of multiple models across various parameters, including solar penetration levels, dataset resolutions, and different hours of the day, enhancing stakeholders' confidence in model performance.

The aim of our application is to build trust through model comparison, primarily comparing our net load forecasting model with a reference model. We enhanced our model to process inputs at varying resolutions and then devised the reference model, which generates predictions by averaging net load ground truths for the last 30 days at the same time point. Despite lacking predictive ML components, this reference model serves as a benchmark in various net load forecasting competitions, including the recent Net Load Forecasting Prize by the National Renewable Energy Laboratory (NREL) and the U.S. Department of Energy Solar Technologies Office (SETO)~\cite{NetLoadForecastingPrize}. Through the web interface of \systemname, we were able to uncover patterns in the performance that help improve trust in the model outcomes. In this work, we identify the visual analytic tasks that are required to compare these models. These tasks can be extended to compare other similar net load forecasting models, enabling users to make well-informed decisions based on the outcomes of these models.

While previous research has predominantly focused on developing interfaces during the model development phase, tailored to assist model developers, our emphasis lies in the post-hoc evaluation of model performance, specifically addressing the needs of energy scientists and grid operators~\cite{lucas2021lumen, klaus2023visual}. Other studies have explored the performance of probabilistic net load forecasting models through different visualization charts but have often lacked an integrated interactive interface~\cite{wang2017data, henriksen2022electrical}. In contrast, our interactive visual analytic tool provides carefully designed visual cues for comparing model performance across different factors like dataset resolutions and solar penetration levels, thus offering a novel approach in this domain.

In this paper, we first introduce the different models used in this work and the rationale behind choosing them. Subsequently, we outline the identified visual analytic tasks and detail the design decisions guiding the development of our interactive interface. This is followed by some of the observations made by our power scientist collaborator through \systemname that demonstrate its efficacy in comparing model performance across multiple facets. Finally, we conclude by sharing insights gained from this development and discussing some of the future research opportunities in this domain. Additionally, a short demonstration video is available \href{https://bit.ly/trustvis}{here}.

\section{Model description}

We start with a deep learning-based probabilistic model tailored for net load forecasting in high behind-the-meter solar scenarios~\cite{sen2022kpf}. This model has three key components: a kernelized probabilistic forecasting (kPF) module, an autoencoder (AE), and a long short-term memory (LSTM) network. This model effectively captured complex temporal dependencies and uncertainties inherent in net load data, which is crucial for reliable forecasting in environments with high solar penetration. By incorporating kernel methods into probabilistic forecasting, the model handled non-linear relationships and captured subtle variations in net load influenced by solar energy fluctuations. The autoencoder component enhanced feature extraction and dimensionality reduction, facilitating the LSTM network's ability to capture long-term dependencies and predict future net load values accurately. Experimental results demonstrated the superior performance of this model compared to traditional forecasting models, showcasing its efficacy in addressing the challenges posed by high solar scenarios and advancing the state-of-the-art in net-load forecasting methodologies. 

This model was used in the Net Load Forecasting Prize competition hosted by NREL and SETO~\cite{NetLoadForecastingPrize}. However, during this competition, we observed its underperformance on the data provided by the organizers, which had lower resolutions. We discovered that while the autoencoder component excelled with high-resolution datasets (such as 15-minute intervals), it struggled to effectively capture temporal dependencies in lower-resolution datasets (e.g., 1-hour intervals provided by the organizers). Consequently, this limitation led to subpar outcomes generated by the LSTM component. In light of this, we opted to remove the kPF and autoencoder components and instead developed a version of the model solely utilizing the LSTM component. Additionally, fine-tuning the number of layers in the LSTM component yielded significantly improved results in the competition.

The competition employed a reference model to assess model performance across various probability levels. As previously mentioned, this reference model simply utilizes historical input data from the past 30 days to generate probabilistic forecasts for a specific time point. This model can serve as the initial benchmark for assessing the effectiveness of other models. Therefore, in this work, we developed a reference model following the same principles. Since these forecasts are probabilistic in nature, we calculated the Continuous Ranked Probability Score (CRPS) for both the reference model and our model. Subsequently, we computed the Continuous Ranked Probability Skill Score (CRPSS) based on these CRPS scores, evaluating whether our forecast presents an improvement or deterioration compared to the reference forecast~\cite{wilks2011forecast, delsole2016forecast}. A positive CRPSS indicates that the forecast outperforms the reference forecast, whereas a negative value suggests inferior performance. Utilizing these CRPSS values, we compare the performance of our model against the reference model across multiple dates throughout the year, and present these values in \systemname. 

\section{Visual analytics-based design}\label{section:design}
\begin{figure*}
\begin{center}
\includegraphics[width=\textwidth]{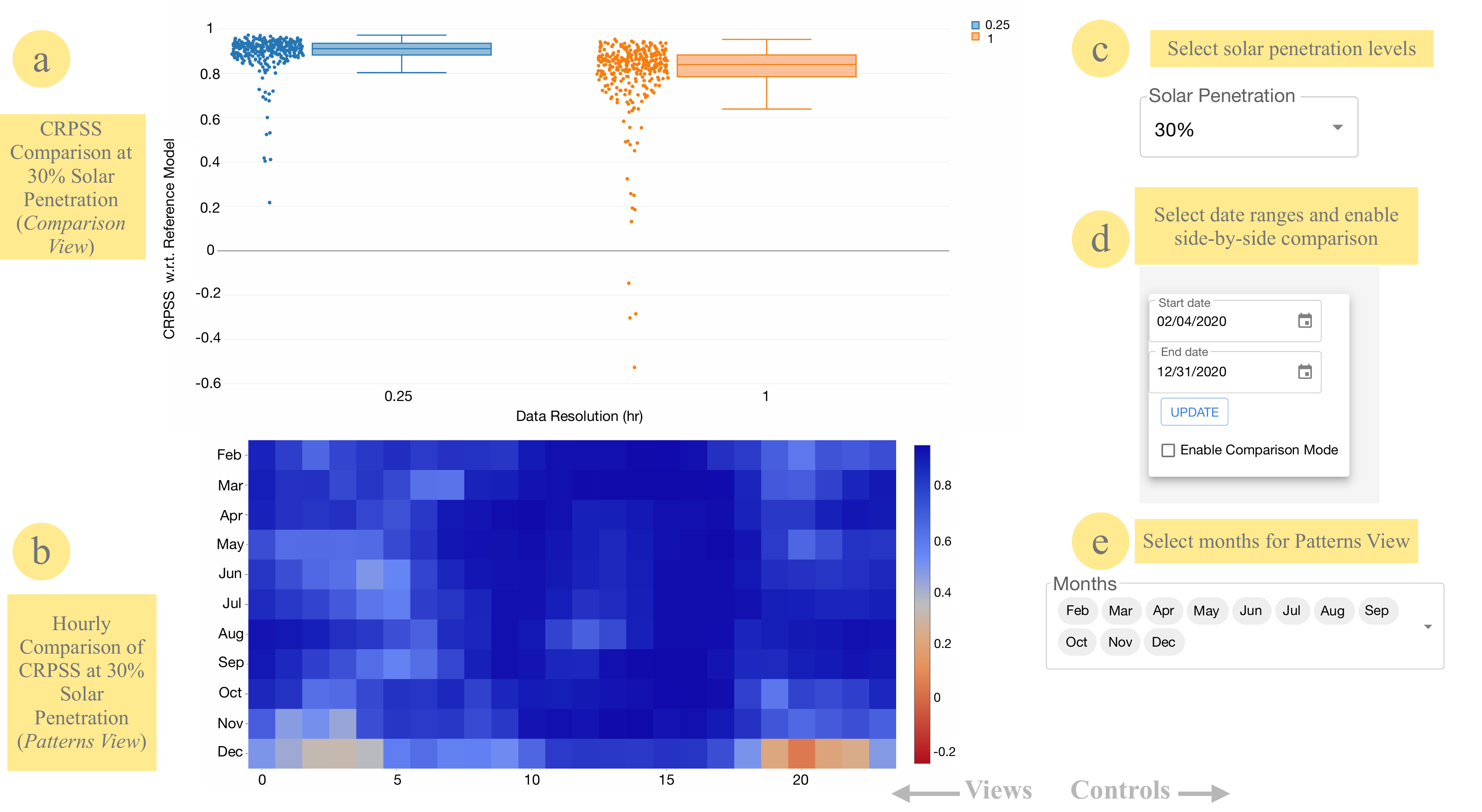}
 \caption{\label{figs:interface}\textbf{Visual analytic application} (a) The Comparison View the facilitates comparison of CRPSS values between the net load forecasting model and the reference model at various data resolutions throughout the year. (b) The Patterns View aids in identifying performance trends across different hours of the day and months. (c), (d) and (e) denote filters for selecting different solar penetration levels, start and end dates, and specific months for the heatmap, respectively.}
\end{center}
\end{figure*}
Our application employs a visual analytics-based design featuring coordinated views enhanced with visual cues to help users compare model performance. It combines interactive visualization with comparison metrics like CRPSS to improve trust in the model outcomes and also allows the users to probe the model and understand its performance across different solar penetration levels and months. In this section, we highlight the two tasks performed by \systemname and how its visual analytics-based design aids in executing these tasks:
\par \noindent \textit{\textbf{T1}: Compare model performance across different solar penetration levels and data resolutions:} Model performance may vary across different solar penetration levels due to the increased variability in net load data caused by intermittent solar generation. LSTM-based models might struggle to capture and predict these dynamic behaviors accurately. Furthermore, datasets with higher resolutions, such as sub-hourly intervals, enable models to more effectively capture short-term fluctuations and dependencies. Hence, this task essentially involves comparing model performance at different solar penetration levels and data resolutions.
\par \noindent \textit{\textbf{T2}: Identify patterns across different timeframes:} By analyzing performance across multiple timeframes, power scientists can assess the net load forecasting models' robustness and consistency in capturing both short-term fluctuations and long-term trends. This evaluation helps to identify whether a model's performance is consistent across various temporal scales or exhibits variability or biases at specific time periods. This task relates to identifying patterns in the model's performance across different months, various hours of the day, and different time periods.

Our application implements multiple coordinated views and components in order to fulfill these tasks. Next, we discuss the design of these views and components along with the rationale behind them:
\par \noindent \textit{\textbf{Comparison View}:} As comparing model performance is the main objective of \systemname, we begin with the Comparison View, which utilizes modified box plots to compare models. Figure~\ref{figs:interface}a depicts CRPSS values on the y-axis and different data resolutions on the x-axis. The box plots illustrate the distribution of CRPSS values, with the median typically exceeding zero, indicating superior performance of our model over the reference in most cases. Users can utilize the solar penetration level filter to compare performance across various levels (20\%, 30\%, 50\%)~(Figure~\ref{figs:interface}c)~\textbf{(T1)}. However, based on initial feedback from domain experts, we recognized the importance of displaying the distribution of CRPSS values. Therefore, we integrated dots representing the CRPSS values, jittered along the x-axis to illustrate their distribution over the entire year. These dots align with insights from the box plot and additionally reveal instances where our model performs worse than the reference on certain dates. This led to the development of the Patterns View, where we can identify the timeframes where the model underperforms compared to the reference.
\begin{figure*}
\begin{center}
\includegraphics[width=\textwidth]{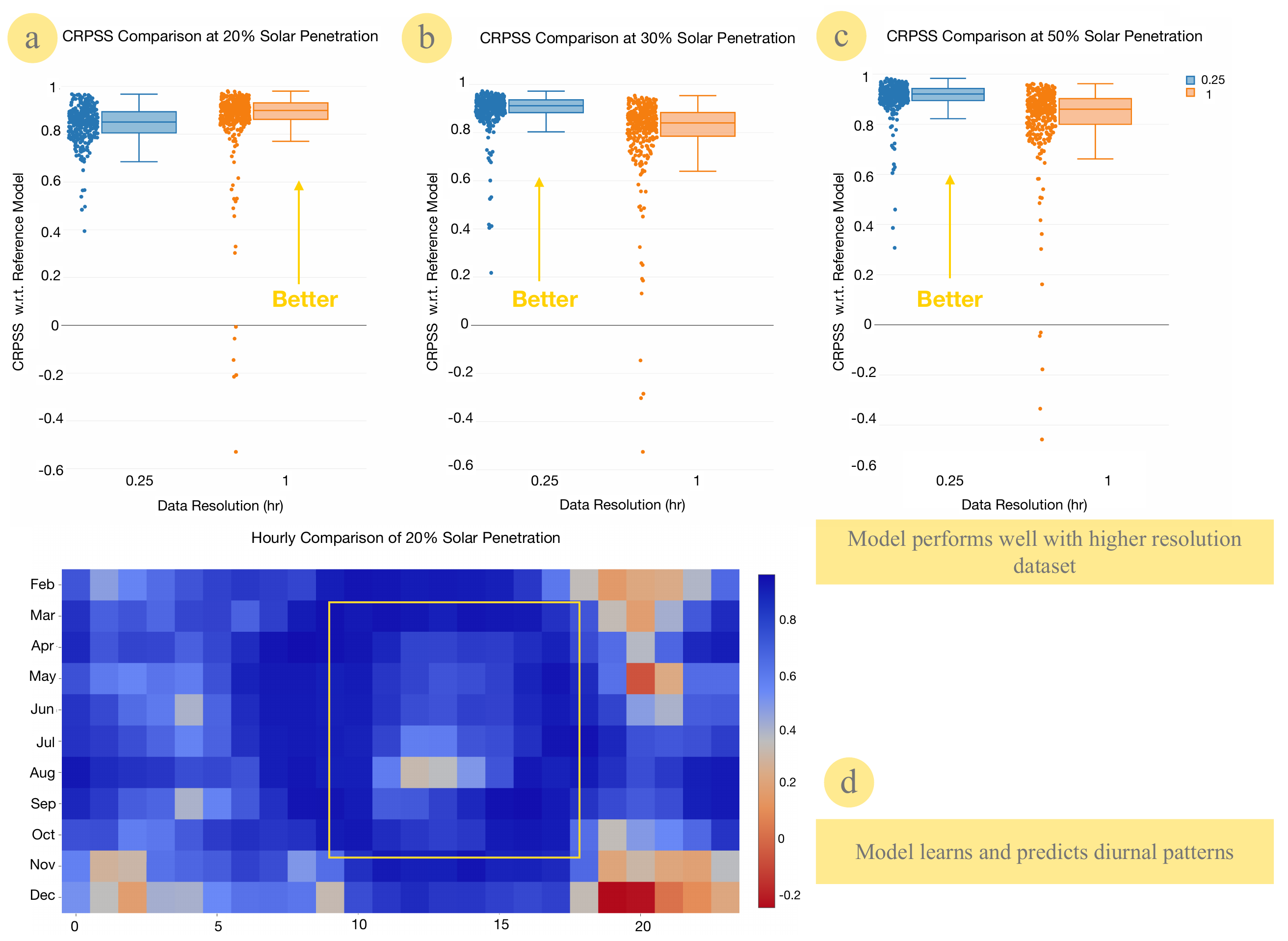}
 \caption{\label{figs:results}\textbf{Results from a case study:} (a), (b), (c) display CRPSS values at varying solar penetration levels, highlighting the model's superior performance with higher-resolution datasets. (d) Additionally, our application reveals insights such as the model's ability to learn and predict diurnal patterns, as evidenced by highlighted box-like patterns.}
\end{center}
\end{figure*}
\par \noindent \textit{\textbf{Patterns View}:} In this view, \systemname utilizes a heatmap to depict performance patterns for each month across different hours of the day~\textbf{(T2)}. The x-axis represents the 24 hours of the day (0-23), while the y-axis displays the months of the year (Feb - Dec)~(Figure~\ref{figs:interface}b). Each box in the heatmap denotes the average CRPSS value for each month at each hour, indicated by the color. Darker blues signify more positive CRPSS values, indicating the superior performance of our model compared to the reference at that time. Conversely, darker reds indicate more negative CRPSS values, signifying poorer performance of our model compared to the reference at that time. This diverging color scale aids in easily identifying performance patterns across different time points and facilitates the identification of instances where our model does not outperform the reference. Users can filter specific months to focus on particular time periods and analyze performance accordingly~(Figure~\ref{figs:interface}e). Additionally, based on feedback from domain experts, we implemented the Sidebar component to allow users to select specific date ranges and analyze performance patterns within those ranges.
\par \noindent \textit{\textbf{Sidebar}:} The Sidebar facilitates user selection of start and end dates to filter results across all views, enabling focus on specific date ranges~\textbf{(T2)} for comparing model performance within those periods~(Figure~\ref{figs:interface}d). Additionally, based on feedback from domain experts expressing interest in comparing model performance across all solar penetration levels simultaneously, \systemname offers a comparison mode toggle in the Sidebar. Enabling this mode updates both views to display box plots and heatmaps for all solar penetration levels side by side, aiding in point-to-point comparison and identification of performance patterns across different solar penetration levels.

\section{Results from a Case Study}\label{section:results}
The efficacy of an application can be validated if the application is able to perform the intended tasks effectively. In this section, we show the results through a case study that demonstrates how \systemname can be used to compare and select net load forecasting models effectively.

This case study involved a power scientist with over $10$ years of experience in power and grid systems. With expertise in nonlinear dynamics, large-scale networks, and distributed control, he played a crucial role in developing the model. His main objective was to assess the model's performance relative to the reference model across various time points and solar penetration levels, which are critical factors to consider before deploying it for a project. We informed him that we had integrated CRPSS values for both the model and the reference model for all dates throughout the year. He then accessed \systemname through a browser and examined the distribution of CRPSS values across different data frequencies at a 20\% solar penetration level~(Figure~\ref{figs:results}a). Notably, he discovered that the model performed better with the lower resolution dataset (1-hour), contrary to expectations for an LSTM-based model~\textbf{(T1)}. Upon further examination, he noted a marginal difference in the median CRPSS between high and low-resolution datasets (0.85 and 0.89, respectively). Therefore, he enabled the comparison mode through the Sidebar, enabling him to assess the model's performance as solar penetration increased. He noted that the model consistently performed well with the higher resolution dataset (15-min) across all other solar penetration levels, confirming the notion that our model excels with high-resolution datasets in high solar penetration scenarios~(Figures~\ref{figs:results}b and ~\ref{figs:results}c). Next, he observed that although the median CRPSS was significantly above zero, the minimum value was negative. Upon inspecting the dots adjacent to the box plots in the Comparison View, he observed that while most dots clustered around the median line, there were a few outliers with negative CRPSS values.

Hence, the scientist sought to understand temporal patterns to identify if any specific hour of the day contributed to the negative values. Consequently, he navigated to the Patterns View within \systemname to analyze the CRPSS value distribution in the heatmap at a 20\% solar penetration level~(Figure~\ref{figs:results}d). Consistent with observations from the Comparison View, most heatmap boxes displayed varying shades of blue, indicating superior model performance compared to the reference across most months and hours. On closer look, the scientist noted a box-like pattern in the heatmap, revealing enhanced performance during morning hours (8 am to 10 am) from April to September, followed by a decline during midday (11 am to 4 pm), and another spike in the evening (4 pm to 6 pm) during these months. This pattern suggested that the model effectively captured diurnal variations in net load data and adjusted predictions accordingly~\textbf{(T2)}. While similar box-like patterns were evident across other solar penetration levels, the intensity of model performance varied with increased solar penetration. This insight holds practical significance for model selection during deployment, as it suggests the potential use of different models or model ensembles tailored to distinct times of the day, leveraging their respective performances on diurnal patterns. Thus, the power scientist was satisfied that \systemname could yield valuable insights regarding the model, aiding informed decision-making. A short demonstration video is available at \href{https://bit.ly/trustvis}{this link}.
\section{Conclusion}\label{section:conclusion}
Efficient model selection for net load forecasting plays a pivotal role in energy planning and grid operations. In this work, we delve into this process, integrating visual analytics with input from domain experts to identify key tasks for comparative model selection. Subsequently, we translate these tasks into an interactive interface, enabling users to assess model behavior across various factors such as solar penetration levels, data resolution, and time of day.

Throughout this endeavor, we gained valuable insights into the model behavior and the challenges posed during the multi-way comparison of models. This collaborative effort underscored the need for interactive tools that facilitate the seamless translation of model insights into actionable decisions. We can argue that our application is a first step towards this direction. As a next step, our plan is to incorporate multiple net load forecasting models into the application and integrate additional metrics for effective comparison of their performance.

Looking ahead, we aim to enhance our application in several ways. Incorporating economic planning and analysis options will allow stakeholders to assess the cost-benefit ratio before model selection, thereby enhancing trust in the outcomes. Additionally, as the energy landscape evolves, our application's flexibility will be pivotal in effectively comparing and selecting forecasting models for real-world applications. In summary, our collaborative endeavor has produced a robust tool for multi-faceted model comparison and has paved the way for informed decision-making through visual analytics in energy planning.


\bibliographystyle{abbrv-doi}
\bibliography{bib}

\end{document}